\documentclass[showpacs,aps,prl,twocolumn,nofootinbib]{revtex4}
\usepackage{amsmath}
\usepackage{epsfig}
\usepackage{graphicx}
\usepackage{subfigure}
\usepackage{float}
\usepackage{verbatim} 
\usepackage{pstricks}
\usepackage{color}
\usepackage{slashed}
\usepackage{multirow}

\newcommand{\gmu}{\gamma^\mu}
\newcommand{\ie}{{\it i.e.}}
\newcommand{\eg}{{\it e.g.}}
\newcommand{\lag}{{\cal L}}
\newcommand{\be}{\begin{equation}}
\newcommand{\ee}{\end{equation}}
\newcommand{\nn}{\nonumber}
\def\bsp#1\esp{\begin{split}#1\end{split}}

\begin{document}

\leftline{}
\rightline{CP3-11-21; IPHC-PHENO-11-03}

\title{Monotops at the LHC}

\author{J.~Andrea$^{a}$, B.~Fuks$^{a}$, F.~Maltoni$^{b}$}
\affiliation{
$^{(a)}$Institut Pluridisciplinaire Hubert Curien/D\'epartement Recherches
  Subatomiques, Universit\'e de Strasbourg/CNRS-IN2P3, 23 Rue du Loess, F-67037 Strasbourg, France\\
$^{(b)}$Center for Cosmology, Particle Physics and Phenomenology (CP3), 
Universit\'e Catholique de Louvain, B-1348 Louvain-la-Neuve, Belgium 
}

\begin{abstract}
We explore scenarios where top quarks may be produced singly in association
with missing energy, a very distinctive signature, which, in analogy with
monojets, we dub {\it monotops}. We find that monotops can be produced in a
variety of modes, typically characterized by baryon number-violating  or flavor-changing neutral interactions. We 
build a simplified  model that encompasses all the possible (tree-level) 
production mechanisms and  study the LHC sensitiveness to a few representative scenarios by considering
fully hadronic top decays. We find that constraints on such exotic models can already be set with 
1 fb$^{-1}$ of integrated luminosity collected at $\sqrt{s}=$7 TeV.
\end{abstract}

\pacs{12.38.Bx,12.60.-i,14.65.Ha}

\maketitle

\section{Introduction}

With the LHC running, the TeV scale has just begun to be explored. 
The search, in particular, of new phenomena, interactions and/or particles,
motivated by several theoretical arguments as well as current precision data, 
moves in several directions. The most beaten path is a top-down approach:
a theory is conceived so that it extends the standard model (SM), addresses one
or more open issues (such as the hierarchy problem or neutrino mass generation),
and predictions can be made through perturbation theory or symmetries. 
In general, some or many new parameters enter, they cannot be fixed 
by the present constraints and yet determine the expected signatures at
colliders.

The most famous example is supersymmetry (SUSY), and more especially its minimal
version \cite{Nilles:1983ge, Haber:1984rc}, which, in its general form at
TeV energies, might feature up to a hundred of free parameters.
Benchmark choices are then made that simplify the analyses and typical signatures
can be identified. For example, in SUSY, same sign leptons are typically
connected to the Majorana nature of some of new states, multijets with missing
energy to heavy colored states decaying to partons and to a stable neutral weakly
interacting state (possibly a dark matter candidate), and multiphotons with
missing energy to gravitinos.

While widely used, this theory-driven approach has notable limitations.
The first is that signatures are neither typical of a given benchmark nor
of a model itself. Universal extra dimensions \cite{Appelquist:2000nn} and SUSY can have, for instance, 
very similar signatures. The second and most important limitation is that the
top-down approach can lead to strong biases in the final state signatures 
 studied in the experimental analyses. 

In this paper, we follow an alternative approach, \ie, 
we propose a final state signature, a top quark in association with 
missing transverse energy, dubbed \textit{monotop}, that no process
in the standard model can lead to at tree-level, the dominant
production mode being suppressed both by a loop factor and by two
powers of nondiagonal Cabibbo-Kobayashi-Maskawa matrix elements.
We are inspired by the monojet and missing energy final state
used to probe gravitons, where the missing energy can be related to weakly
interacting states leaving no visible traces in the detectors.
Having a top quark in the final state, however, gives a signature much clearer
and easier to discriminate than just a light jet. From the theoretical
point of view, asking for a top gives a further advantage in that it 
fixes the flavor of the final state and restricts the possibilities
for partons in the initial state. As a result, we find that there are only two 
classes of processes leading to such a final state signature,
through baryon number-violating and flavor-changing neutral interactions.
In both cases the key assumption  is the existence of an enhanced coupling between the first and third
generation. Interestingly enough, both these phenomena are not very much
constrained for the top quark~\cite{Nakamura:2010zzi} and room appears for the LHC 
to make a discovery or set bounds.

\section{Beyond the Standard Model Explorations}

\begin{figure}[t]
  \includegraphics[width=.49\columnwidth]{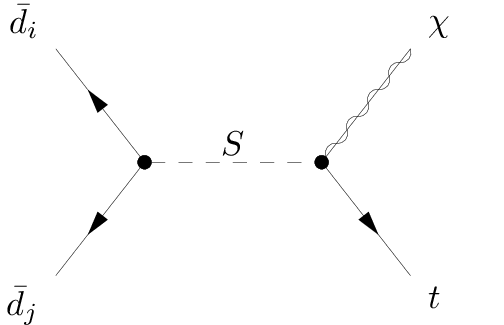}
  \includegraphics[width=.49\columnwidth]{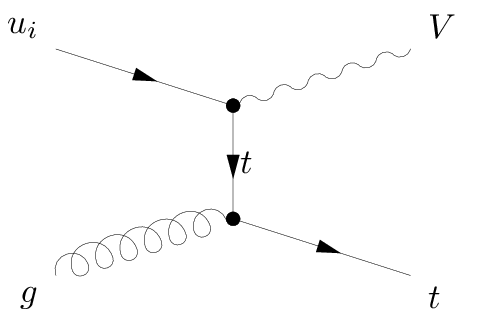}
  \caption{\label{fig:01} Representative Feynman diagrams leading to monotop
    signatures, through the resonant exchange of a colored
    scalar field $S$ (left) and via a flavor-changing 
    interaction with a vector field $V$ (right). In these two examples, the
    missing energy is carried by the $V$ and $\chi$ particles. More diagrams
    with, for example, $t$-channel and $s$-channel exchanges for the two
    type of processes respectively, are possible.}
\end{figure}

At the tree level, monotop production can occur via two main mechanisms. 
Either the top quark is produced (resonantly or not) in association with
missing energy of a fermionic nature or through a flavor-changing interaction with an
invisible bosonic state, as illustrated on the left and right panels of Fig.\
\ref{fig:01}, respectively.

In the first class of models, the top quark is produced in association
with an undetected fermionic particle $\chi$ via $s,t,u$-channel exchanges of a scalar ($S$) or vector ($V$) field lying in the
(anti-) fundamental representation of $SU(3)_c$.  As an example, consider 
the $s$-channel resonant case
\be 
  \bar d_i \bar d_j  \to  S\ \text{or}\ V  \to  t \chi\ ,
\nn\ee
where $d_k$ denotes a down-type quark of generation $k$. 
Such processes occur in $R$-parity-violating SUSY \cite{Barbier:2004ez} where,
similarly to the case discussed in Ref.\ \cite{Desai:2010sq}, the intermediate
particle is a (possibly on shell) squark and $\chi$ the lightest neutralino ($\bar
d \bar s \to \tilde u_i \to t \tilde \chi_1^0 $, where $\tilde u_i$ are any of the up squarks), or in $SU(5)$ theories
where a vector leptoquark $V$ decays into a top quark and a neutrino ($\bar d
\bar d \to V \to t \bar \nu$). The key difference between these two examples is the
mass of the invisible fermionic state inducing different 
transverse-momentum ($p_T$) spectrum for the top quark. In the limit of a
very heavy resonance, monotops can be seen as being produced through a baryon
number-violating
effective interaction ($\bar d \bar s \to
t \bar \chi$), after having included the possible $t$- and $u$-channel exchanges of
a heavy field \cite{Morrissey:2005uza, Dong:2011rh}. Let us note that the fermionic particle could also be a Rarita-Schwinger field,
as in SUSY theories containing a spin-$3/2$ gravitino field, or a multiparticle
state (with a global half-integer spin), as in hylogenesis scenarios for dark
matter \cite{Davoudiasl:2011fj}.
 
In the second class of models, the top quark is produced in association with a
neutral bosonic state, either long-lived or decaying invisibly, from
quark-gluon initial states undergoing a flavor-changing
interaction, as discussed, \eg, in Ref.\ \cite{delAguila:1999ac}. Missing energy consists either in a two-fermion continuous state,
as in $R$-parity conserving SUSY \cite{Allanach:2010pp}, or in a spin-0 ($S$),
spin-1 ($V$) or spin-2 ($G$) particle,
\be
 u g \to \tilde u_i \tilde\chi_1^0 \to t \tilde \chi_1^0 \tilde \chi_1^0 \ , \quad
 u g \to t S \ , \ t V \  \text{ or }\   t G \,.
\nn\ee

\section{Effective theory for monotops}

The top quark kinematic distributions depend both on the partonic
initial state and on the nature of the undetected recoiling object
(scalar, massive or massless fermion, vector or tensor), as well as on the
possible presence of an intermediate resonant state. This suggests a 
model-independent analysis where we account for all cases within a single
simplified theory, in the same spirit as Ref.~\cite{Alves:2011wf}. 
Assuming QCD interactions to be  flavor-conserving, as in the SM, the flavor-changing neutral interactions are
coming from the weak sector. We denote by $\phi$, $\chi$ and $V$ the possible scalar, fermionic and vectorial
missing energy particles, respectively and by  $\varphi$ and $X$ scalar and
vector fields lying in the fundamental representation of $SU(3)_c$ which could
lead to resonant monotop production.\footnote{For simplicity, we neglect
spin-2 gravitons, as their flavor-changing couplings are
loop-induced and thus very small \cite{Degrassi:2008mw}, as well as any of their
excitations, which, even if they have, on the one hand, typically
flavor-violating couplings at tree level, do not lead, on the other hand, to a
missing energy signature.
On the same footing, we do not consider spin-$3/2$ fields since their couplings
are, at least in SUSY theories, in general suppressed by the SUSY-breaking scale.}
In addition, we obtain a simplified modeling of four-fermion interactions through
possible $s,t,u$ exchanges of heavy scalar fields $\varphi$ and $\tilde \varphi$. The corresponding
effective Lagrangian in terms of mass eigenstates reads
\be\label{eq:lag}\bsp
 \lag &=\ \lag_{SM} \\
  &+ \phi \bar u \Big[a^0_{FC}\!+\!b^0_{FC} \gamma_5 \Big] u \!+\!
     V_\mu \bar u \Big[a^1_{FC} \gmu \!+\! b^1_{FC} \gmu \gamma_5 \Big] u  \\
  &+\! \epsilon^{ijk} \varphi_i \bar d^c_j 
       \Big[a^q_{SR} \!+\! b^q_{SR} \gamma_5 \Big] d_k \!+\!
     \varphi_i \bar u^i \Big[a^{1/2}_{SR} \!+\! b^{1/2}_{SR} \gamma_5 \Big] \chi \\
  &+ \! \epsilon^{ijk} \tilde\varphi_i \bar d^c_j 
       \Big[\tilde a^q_{SR} \!+\! \tilde b^q_{SR} \gamma_5 \Big] u_k \!+\!
     \tilde \varphi_i \bar d^i \Big[\tilde a^{1/2}_{SR} \!+\! \tilde b^{1/2}_{SR} \gamma_5 \Big] \chi \\
   &+ \epsilon^{ijk} X_{\mu,i}\ \bar d^c_j 
       \Big[a^q_{VR}\gmu + b^q_{VR} \gmu\gamma_5 \Big] d_k  \\
   &+ X_{\mu,i}\ \bar u^i 
       \Big[a^{1/2}_{VR} \gmu + b^{1/2}_{VR} \gmu \gamma_5 \Big] \chi + 
       {\rm h.c.}  ,
\esp\ee
where the superscript $c$ stands for charge
conjugation, $i,j,k$ are color indices in the fundamental representation and 
flavor indices are understood. The matrices (in flavor space) $a^{\{0,1\}}_{FC}$
and $b^{\{0,1\}}_{FC}$ contain quark interactions with the bosonic
missing-energy particles $\phi$ and $V$, while $a^{1/2}_{\{S,V\}R}$ and
$b^{1/2}_{\{S,V\}R}$ are the interactions between up-type quarks,
the invisible fermion $\chi$ and the new colored states $\varphi$ and $X$. The
latter also couple to down-type quarks with a strength given by $a^q_{\{S,V\}R}$
and $b^q_{\{S,V\}R}$. Because of the symmetry properties of the $\epsilon^{ijk}$
tensor, identical quark couplings to the scalar field $\varphi$ vanish and so do
their axial couplings to the vector field $X$. In the case of
four-fermion interactions, we also need to introduce additional $\tilde
a^q_{SR}$, $\tilde b^q_{SR}$, $\tilde a^{1/2}_{SR}$ and $\tilde b^{1/2}_{SR}$
interaction matrices, assuming heavy masses for the $\varphi$ and
$\tilde\varphi$ fields.

\section{Model-independent searches}

The main signatures associated with monotop production can be classified according
to the top quark decays,
\be
  p p \to t + \slashed{E}_T \to b W + \slashed{E}_T \to b j j + \slashed{E}_T
    \quad \text{or} \quad b \ell + \slashed{E}_T \ ,
\nn \ee
where $j$ and $b$ denote (parton-level) light/$c$- and $b$-jets, respectively, $\ell$
a charged lepton and $\slashed{E}_T$ missing transverse energy. In this paper,
we  consider the simpler hadronic signatures of one $b$-jet and two light
jets, typically lying in the same hemisphere. While monotop signatures with
leptonically decaying top quarks are expected to be more challenging, they have
been widely investigated in the
past in the context of $R$-parity-violating SUSY \cite{Berger:1999zt, Berger:2000zk}. Top quark mass reconstruction is a 
powerful tool to reject electroweak or QCD backgrounds at low as well as at high transverse momentum. In the latter case, boosted top reconstruction techniques could be also exploited to reconstruct fully hadronic top candidates with a good purity \cite{Plehn1}.

The only source of irreducible SM background to hadronic monotop production
consists in the associated production of an invisibly decaying $Z$-boson with
three jets (one being a $b$ jet). Other sources of background, related to
detector effects, consist, first, in QCD multijet events where misreconstructed jets
produce large (fake) $\slashed{E}_T$, then in $W$ plus jets, $t\bar{t}$ and
diboson events where the $W$ and $Z$ bosons decay to nonreconstructed leptons,
and finally in single top events including non- or misrecontructed jets. A proper
investigation of those instrumental backgrounds requires not only parton showering,
hadronization, and realistic detector simulation, but also data-driven methods. 
It is therefore preferable to resort to the available experimental studies, \eg, \cite{CMSpaper,daCosta:2011qk}. 
In a recent analysis performed by the CMS collaboration with the 7 TeV data~\cite{CMSpaper}, 
it has been shown that simple selection cuts on the missing transverse momentum  $(>150$ GeV) and 
on the $p_T$ $(>50$ GeV) of the three jets reconstructed with a high quality as
well as on their scalar sum ($> 300$ GeV) allow to keep a good control of the backgrounds.
This leads to comparable amounts of selected QCD, $Z$, $W$ and
$t\bar{t}$ events, while the contamination from diboson and single top
backgrounds is further reduced by about one order of magnitude. In addition, 
the presence of a top quark can be exploited by
demanding two non-$b$-jets  with an invariant mass  compatible with the
$W$-boson mass, one $b$-tagged jet, and a three-jet invariant mass compatible
with the top quark mass. Hence, all instrumental backgrounds are expected to
be strongly suppressed after selection. We therefore base our estimate of the monotop sensitivity at the LHC
taking into account the  irreducible background related to the production of an invisibly decaying 
$Z$-boson in association with three jets.

Event simulation is performed for the LHC at $\sqrt{s} = 7$ TeV and for a
luminosity of 1 fb$^{-1}$ using the Monte Carlo generator {\sc MadGraph} 5~\cite{Alwall:2011uj}. 
We employ the CTEQ6L1 set of parton densities~\cite{Pumplin:2002vw}
and identify renormalization and factorization scales to the value of
the top quark mass, $M_t = 172$ GeV for the signal (as well as for the background). 
We have implemented the Lagrangian of Eq.~\eqref{eq:lag} into {\sc MadGraph} via  
{\sc FeynRules}  \cite{Christensen:2008py}.   In this prospective
study, we present results based on a parton-level simulation for the signal 
as well as for the main irreducible background, $Z+3$ jets,  computed at the tree level in the five-flavor scheme.

To illustrate  the main features of the monotop production, we consider
simplified scenarios where all axial couplings involving new particles vanish
($b \!=\! \tilde b \!=\! 0$). Furthermore, we
only retain  the interactions that can be enhanced by parton density functions, 
and set $(a^0_{FC})_{13} = (a^0_{FC})_{31} = 
    (a^1_{FC})_{13} = (a^1_{FC})_{31} =
    (a^q_{SR})_{12} = \ -(a^q_{SR})_{21} = (a^{1/2}_{SR})_3 =
    (a^q_{VR})_{11} = (a^{1/2}_{VR})_3 = a = 0.1$.
In the four-fermion interaction limit, the heavy mass
is set to $M_\varphi \!=\! M_{\tilde\varphi} \!=\! 3$ TeV and the corresponding
nonvanishing interaction are $(a^q_{SR})_{12} = -(a^q_{SR})_{21} =  (\tilde a^q_{SR})_{\{1,2\}3} = (\tilde
 a^{1/2}_{SR})_{\{1,2\}} = a =0.1$\ .
Within the above settings, we define five scenarios:  a scalar (vector) resonant monotop
production with $m_\chi \!=\! 50$ (300) GeV,  flavor changing neutral production 
with a 300 GeV scalar (50 GeV vector) invisible state and, finally, a scenario including
four-fermion interactions with a massless invisible state $\chi$. 
The values for the masses of the invisible states are inspired by present collider 
data and lie right above the lower bound on the mass of the lightest neutralino in typical
SUSY scenarios  and of the lightest Kaluza-Klein
excitation in extra-dimensional scenarios, respectively~\cite{Nakamura:2010zzi}. In the case of resonant monotop production, we
assume the branching ratio $X,\varphi \to t \chi$  equal to one, our
analysis being thus insensitive to $a^q_{\{SR,VR\}}$, and we fix
the resonance mass to 500 GeV. 
\begin{figure}[t!]
  \centering
  \includegraphics[width=\columnwidth]{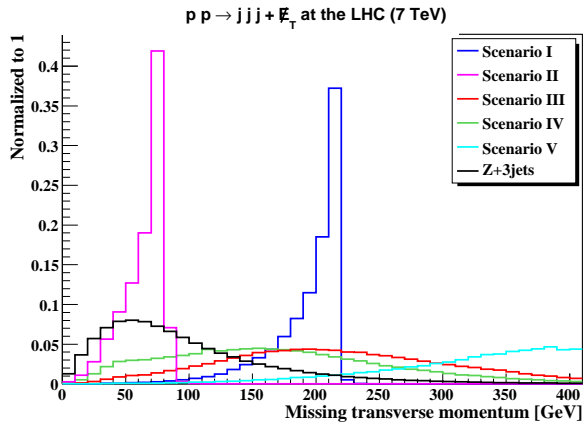}
  \caption{Missing transverse-momentum spectrum for monotop
     production ($t + \bar t$) at the LHC (7 TeV) for the five considered scenarios
     and for the $Z\!+\!3$ jet background. Selection cuts are given in the text. 
     The distributions are normalized to unity. }
  \label{fig:03}
\end{figure}
In Fig.\ \ref{fig:03}, we present the missing transverse-momentum
($\slashed{p}_T$) spectrum associated with the five monotop scenarios
presented above and with the irreducible $Z\!+\!3$ jet background. 
We require three jets with $p_T>50$ GeV, rapidity $|\eta|<2.5$, and relative distance $\Delta R>0.5$. 
All event samples being normalized to one, we compare
the shapes of the distributions. As expected, we observe a typical resonant behavior for the first two
scenarios, the spectrum showing an edge at a value depending on the masses of
the resonance and of the invisible produced fermion.
On the other hand, the distributions related to flavor-changing monotop production
are flatter, peaking at higher $\slashed{p}_T$ values with respect to the background. 
Finally, monotop production through four-fermion interactions leads to a $\slashed{p}_T$-distribution
monotonically growing with the energy, thus fully distinguishable from the SM
background if the number of signal events is large enough.

\begin{table}[t]
 \caption{\label{tab:01}Signal cross sections in the five selected scenarios 
 corresponding to the choice $a=0.1$.  In scenarios I  and II the  branching ratios of intermediate resonances is set to unity. In the last column the minimal values of the effective couplings   $a_{min}$ leading to sensitivities $s=S/\sqrt{S+B}\ge 5$, 
 assuming 1 fb$^{-1}$ of collected luminosity at LHC ($\sqrt{s}=7$ TeV) are given.}
 \begin{tabular}{|c | c c  c | c |}
  \hline Scenario & $\slashed{p}_T$ cut [GeV] & $\sigma(t +\bar t)$ [pb]&  $\ $ $a_{min}$ \\
  \hline
   I & 150 & 3.99  & 0.042\\ 
   II & 65  & 32.1 & 0.043\\
   III & 150 & 0.322& 0.14\\
   IV & 150 & 24.3& 0.017\\
   V & 250 & 1.08$\cdot 10^{-4}$& 4.9\\
   \hline
 \end{tabular} 
\end{table}

In summary, $\slashed{p}_T$ distributions are characteristic of  the 
monotop production mode. However, in all cases, simple cuts on the missing transverse momentum
allow us to get rid of the major background contributions and keep a good
fraction of the signal. We define three basic possible cuts on the required
$\slashed{p}_T$: ``loose'', ``standard'' and ``tight'' cut corresponding to $\slashed{p}_T > 65, 150, 250$ GeV
respectively. The first is suitable for resonant monotop production when the resonance mass is close to the
monotop production threshold. The latter leaves an almost background-free sample, useful 
in the production via four-fermion interactions. For all intermediate cases we employ the standard cut.
 Our leading-order estimate of  the cross section for the $Z$ (invisible) + 3 jet background within the jet cuts mentioned above and the additional $\slashed{p}_T>65$ GeV cut  gives 4.9 pb.
 In the context of an analysis including parton showering, hadronization and detector simulation, as well as all
instrumental backgrounds,  events containing isolated leptons should be rejected and exactly three jets
required. In so doing, background from single top and $t\bar t$ events becomes negligible and  combinatorial issues for the top candidate reconstruction are minimized, at a price of a  lower signal selection efficiency. 
In more refined analysis, the $p_T$-threshold could be further tuned
to optimize over extra radiation possibly present in the events. We only retain events with a 
single $b$-tagged jet, estimating a $b$-tagging efficiency of about 60\% for a charm/light flavor 
mistagging rate of 10\%/1\%. In addition,  the two non-$b$-tagged jets invariant-mass is required 
to lie in a  $M_W \pm 20$ GeV range while the three-jet invariant-mass is required to be in a $M_t \pm 30$ GeV interval. 
In order to include resolution effects, the (parton-level) invariant-mass
distributions have been smeared using Gaussian functions with a width of 15 GeV in
the dijet and 20 GeV in the three-jet cases. 

\begin{figure}[t!]
  \centering
  \includegraphics[width=\columnwidth]{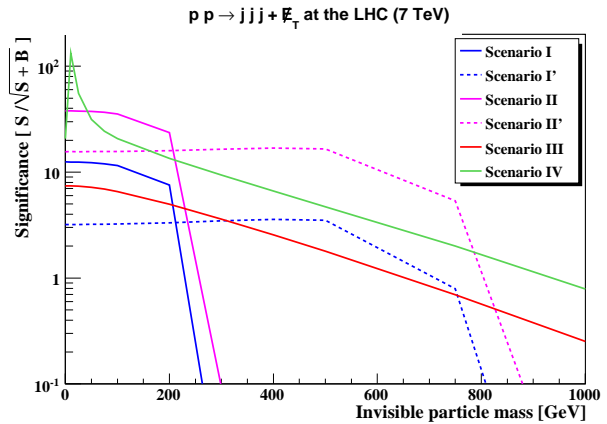}
  \caption{Significance in the five selected scenarios as a function of the invisible state mass, assuming 1 fb$^{-1}$ of collected luminosity at LHC ($\sqrt{s}=7$ TeV). Selection cuts are described in the text.}
  \label{fig:04}
\end{figure}

In Table \ref{tab:01} the signal cross sections and the corresponding sensitivities at LHC $\sqrt{s}=7$ TeV with 
1 fb$^{-1}$ accumulated luminosity are shown for the five representative scenarios. 
In Fig.\ \ref{fig:04}, we estimate the sensitivity of the LHC to  monotop discovery. 
Starting from the five scenarios presented above, we
vary the mass of the missing energy particle and calculate in each case the
significance, using the standard $\slashed{p}_T$ cut of 150
GeV. The benchmark scenarios I' and II' are variations of the two first
scenarios where the mass of the resonant particle is taken at 1 TeV.
Flavor changing monotop production appears to be the most optimistic case,
especially in the low mass region, due to the enhancement  given by the parton
densities. On the other hand, in the case
of resonant monotop production, the accessible mass region depends strongly on
the resonance mass, with heavier resonances allowing to probe a larger invisible
mass.

\section{Conclusions}
In this paper, we have proposed a novel signature, which we dubbed monotop, involving the 
production of a top in association with missing energy. We have analyzed it 
through a simplified theory approach and argued that couplings of order one or smaller 
can be strongly constrained at the LHC  for the different production modes, just by using a basic set of cuts. 
Our results motivate further studies, possibly including more complete simulations and advanced analysis techniques, \eg, the use of boosted top reconstruction algorithms, as well as  a thorough analysis of the indirect constraints coming
from lower energy experiments and flavor physics. 

\section*{Acknowledgments}
The authors are grateful to J.~M.~G\'erard for valuable discussions and a careful
reading of the manuscript. B.~F.\ acknowledges support by the Theory-LHC
France-initiative of the CNRS/IN2P3. F.~M.\ is supported by the Belgian IAP
Program, BELSPO P6/11-P and the IISN convention 4.4511.10.

\bibliography{review}

\end{document}